\documentclass[preprint,aps,showpacs,draft]{revtex4}
\usepackage{graphicx}
\usepackage{dcolumn}
\usepackage{bm}
\usepackage{epsf}
\usepackage{color}
\usepackage{ulem} 

\newcommand{\edc}{\end{document}}
\newcommand{\bb} {\begin{thebibliography}}
\newcommand{\eb}{\end{thebibliography}}
\newcommand{\bi}[1]{\bibitem{#1}}
\newcommand{\bc}{\begin{center}}
\newcommand{\ec}{\end{center}}

\newcommand{\be}{\begin{equation}}
\newcommand{\ee}{\end{equation}\normalsize}
\newcommand{\bea}{\begin{eqnarray}}
\newcommand{\eea}{\end{eqnarray}}
\newcommand{\ba}{\begin{array}{l}   }
\newcommand{\lab}[1]{\label{#1}}
\newcommand{\ea}{\end{array}}

\newcommand{\dsfrac}{\displaystyle\frac}
\newcommand{\ds} {\displaystyle}
\newcommand{\summa}{\ds\sum}
\newcommand{\re}[1]{(\ref{#1})}

\newcommand{\ci}{\cite}

\newcommand{\akrest}{{a}^\dagger}

\newcommand{\bkrest}{{b}^\dagger}

\newcommand{\dsint}{\ds\int}

\newcommand{{\vergul}}{  ,}

\newcommand{\veps}{\varepsilon }

\newcommand{\rhozero}{\ds{\rho_0}}
\newcommand{\rhoone}{\ds{\rho_1}}

\newcommand{\sigmaone}{\ds{\sigma_1}}

\begin{document}
\title{High-field instability of field-induced triplon Bose-Einstein condensate}
\author{Abdulla Rakhimov$^{a}$} 
\author{E. Ya. Sherman $^{b,c}$} 
\author{Chul  Koo Kim$^{d}$}

\affiliation{
$^a$Institute of Nuclear Physics, Tashkent 100214, Uzbekistan \\
$^b$ Department of Physical Chemistry, University of Basque Country, 48080 Bilbao, Spain \\
$^c$ IKERBASQUE Basque Foundation for Science, Alameda Urquijo 36-5,
     48011 Bilbao, Bizkaia, Spain \\
$^d$ Institute of Physics and Applied Physics, Yonsei University, Seoul 120-749, R.O. Korea\\
}

\begin{abstract}
We study properties of magnetic field-induced Bose-Einstein condensate
of triplons as a function of temperature and the field
within the Hartree-Fock-Bogoliubov approach including
the anomalous density. We show that the magnetization is continuous 
across the transition, in agreement with
the experiment. In sufficiently strong fields the condensate becomes unstable due to 
triplon-triplon repulsion. As a result, the system is characterized by two critical
magnetic fields: one producing the condensate and the other destroying it. 
We show that nonparabolic triplon dispersion arising due to the gapped bare spectrum 
and the crystal structure has a strong influence on the
phase diagram.
\end{abstract}
\maketitle

Bose-Einstein condensation (BEC), a macroscopic quantum
phenomenon, occurs in various systems of bosons, including, in addition to atoms,
quasiparticles in systems out of equilibrium such as excitons
and polaritons (for example,\cite{Snoke07}). Theory predicts that
quantum spin excitations in solids, 
being Bose-quasiparticles, can at certain conditions build the condensate,
and magnetic ordering in various sysems
can be understood in terms of the BEC of these excitations.\cite{Matsuda,Tachiki,Batyev,Affleck,Giamarchi99}
The first experimental observation \cite{Nikuni00} of magnetic 
field-induced BEC of triplons, that is the
spin $S=1$ quasiparticles, in antiferromagnetic TlCuCl$_3$ produced a 
diverse research field. 
\cite{Crisan05,Stone07,Ruegg07,Yamada08,Amore08,Aczel09,Paduan09,Laflorencie09,Demidov08,Giamarchi08} 
In this compound, the triplon branches
with $S_z=-1,0,1$, are separated
from the ground state by a relatively small gap $\Delta$. For this reason,
the Zeeman interaction in a modest external magnetic field $H_{\rm ext}$ can close
the gap for the $S_z=-1$ states. In contrast to atomic gases,
where the total particle number is constant, for triplons it is
proportional to magnetization $M(T,H_{\rm ext})$ induced by $H_{\rm ext}$.
The density of triplons rapidly increases with the field, 
and they undergo the BEC leading to a magnetic ordering. 
This field-induced BEC, which occurs at the scale on the order of few K,
has been observed in a variety of quantum antiferromagnets.\cite{Giamarchi08}

The condensate properties crucially depend on interaction of the particles.\cite{Huangbook}
For the atomic BEC at $T=0$ the interatomic repulsion can
lead to the condensate instability when the concentration becomes large enough.\cite{Rakhimov08} 
Another general feature clearly seen in the triplon BEC is the dependence of its physics
on the bare dispersion of the quasiparticles $\varepsilon_{\bf k}$. The non-parabolic bare dispersion
of triplons \ci{Misguich04} leads to a non-trivial dependence
of the transition temperature $T_{\rm c}$ on the concentration
$\rho\sim M(T,H_{\rm ext})$ and, hence, on $H_{\rm ext}$. 
The bare dispersion, being itself $H_{\rm ext}$-independent, 
determines the interplay of kinetic and potential energy 
of a macroscopic system, and, therefore, plays a crucial
role in the BEC properties. 
The effects of the bare dispersion are clearly seen experimentally 
as the $\rho$-dependence $T_{\rm c}\sim\rho^{\phi(\rho)}$.
The exponent ${\phi(\rho)}$ approaches $2/3$ at low concentrations (low $T_c$),\ci{Yamada08}
as predicted for the parabolic $\varepsilon_{\bf k}$, while  
at $T>2.5$ K, ${\phi(\rho)}$ is close to 0.5.  

Here we establish theoretical phase diagram of the field-induced triplon BEC
based on the Hartree-Fock-Bogoliubov (HFB) approximation taking into account also a nonparabolic
dispersion and determine the fields $H_{\rm ext}^{(1)}$ and $H_{\rm ext}^{(2)}>H_{\rm ext}^{(1)}$,
corresponding to the BEC onset and to the instability. A problem in the current
theoretical description of the transition at $H_{\rm ext}^{(1)}$
is its predicted discontinuity. 
We show that this result is an artefact of the conventional 
Hartree-Fock-Popov (HFP) approximation, neglecting the anomalous density terms. 
When the anomalous density is taken into account, the theory
correctly predicts the continuous transition. 
For this reason, the HFB method is more appropriate 
to study the BEC than the HFP one. 
We find here the stability region of the triplon BEC in the $T-H_{\rm ext}$
plane and prove that its boundaries strongly depend on the dispersion
$\varepsilon_{\bf k}$.  
Results on triplons and on cold atoms can be compared to foster the understanding of 
the similarities and differences of their BEC.

The triplons form a non-ideal Bose gas \cite{Batyev,Nikuni00,Misguich04} 
with contact repulsive interaction described by the Hamiltonian:
\be
\ba
\hat{H}=\dsint dV
\Big\{
\psi^{\dagger}({\bf r}){\cal K}\psi({\bf r})+
\dsfrac{g}{2}[\psi^{\dagger}({\bf r})\psi({\bf r})]^2
\Big\},
\lab{eq21}
\ea
\ee
where ${\cal K}$ is the kinetic energy operator  and $g$ is the coupling constant, and
we adopt the units $k_{B}=1$, $\hbar=1$, and $V=1$ for the crystal volume. 
Below the critical temperature $T_{\rm c}$
the global gauge symmetry becomes  broken as realized by the
Bogoliubov shift in the field operator:
$\psi({\bf r})=v+\widetilde{\psi}({\bf r})$.
Here the condensate order parameter $v$ and $\widetilde{\psi}({\bf r})$  define
the density of  condensed and  uncondensed particles, respectively:
$\rhozero=v^2$, $\rhoone=\langle{\widetilde{\psi}}^{\dagger}({\bf r}) \widetilde{\psi}({\bf r})\rangle.$
The grand canonical Hamiltonian is: $\hat{H}_{G}=\hat{H}-\mu \rho$,
where $\mu$ is the chemical potential and the total density $\rho=\rho_0+\rho_1$ is
uniquely determined by $H_{\rm ext}$. The density $\rho$ is considered as a dimensionless quantity.
After the Bogoliubov shift one presents the grand Hamiltonian $\hat{H}_{G}$ 
in terms of second quantization operators as 
$\hat{H}_{G}=H_0+H_1+H_2+H_3+H_4$ with:\ci{Andersen04}
\begin{eqnarray}
&&H_0=-\mu\rhozero+\dsfrac{g\rho_{0}^2}{2},\\
&&H_2={\summa}_{\bf k}^{\prime}\left[
 \left(\varepsilon_{\bf k}-\mu+2g\rhozero\right)\akrest_{\bf k} a_{\bf k}+\dsfrac{g\rhozero}{2}
 \left(a_{\bf k} a_{\bf -k}+{\rm h.c.}\right)
\right],\nonumber\\
&&H_4=\dsfrac{g}{2}
{\sum}_{\bf k,p,q}^{\prime}\akrest_{\bf k} \akrest_{\bf p}a_{\bf q}a_{\bf k+p-q},\nonumber
\end{eqnarray}
where the prime shows that zero momentum states are excluded. Similarly defined linear ($H_1$) 
and cubic ($H_3$) terms having zero  mean-field approximation (MFA) expectation values
are omitted.
  
  Now we implement the HFB approximation \ci{yukalov,Andersen04}:
   \be
   \akrest_{\bf k} \akrest_{\bf p}a_{\bf q} a_{\bf m}\rightarrow
   4\akrest_{\bf k} a_{\bf m}\langle  \akrest_{\bf p}  a_{\bf q}  \rangle+
   a_{\bf q} a_{\bf m}\langle  \akrest_{\bf k}  \akrest_{\bf p}  \rangle+
   \akrest_{\bf k} \akrest_{\bf p} \langle    a_{\bf q} a_{\bf m}  \rangle-
   2\rho_{1}^2-\sigma_{1}^2,
   \ee
where $\langle  \akrest_{\bf k}  a_{\bf p}  \rangle=\delta_{{\bf k},{\bf p}}n_{\bf k}$,
$\langle  a_{\bf k}  a_{\bf p}  \rangle=\delta_{{\bf k},{-\bf p}}\sigma_{\bf k}$, $n_{\bf k}$ and
$\sigma_{\bf k}$ are related to the normal $\rhoone=\summa_{\bf k} n_{\bf k}$
 and anomalous $\sigmaone=\summa_{\bf k} \sigma_{\bf k}$  densities.  The grand Hamiltonian
 in this approximation involves only zero $\widetilde{H}_0$ and second order
$\widetilde{H}_2$ contributions in $a_{\bf k}, \akrest_{\bf k}$:
 \be
 \ba
 \widetilde{H}_0=-\mu\rhozero+\dsfrac{g}{2}\left[\rho_{0}^2-2\rho_{1}^2-\sigma_{1}^2\right],\\
 \widetilde{H}_2={\summa}_{\bf k}^{\prime}\left[\omega_{\bf k}\akrest_{\bf k}a_{\bf k}+\dsfrac{X_1}{4}
\left(a_{\bf k} a_{\bf -k}+{\rm h.c.}\right)\right],
\lab{eq234}
 \ea
 \ee
 where $\omega_{\bf k}=\varepsilon_{\bf k}-\mu+2g\rho$ and
 \be
  X_1=2g(\rhozero+\sigmaone).
 \lab{eq24}
 \ee
  It follows from \re{eq234} that for $T>T_{\rm c}$ the $\widetilde{H}_2$ term
  is diagonal and hence, the triplon density is given
 by the same formula as in the widely used HFP approximation
 \be
 \rho(T>T_{\rm c})=\summa_{\bf k} \dsfrac{1}{e^{\omega_{\bf k}}/T-1}\equiv
 \summa_{\bf k} \dsfrac{1}{e ^{(\varepsilon_{\bf k}-\mu_{\rm eff})/T}-1},
 \lab{eq25}
 \ee
 where $\mu_{\rm eff}=\mu-2g\rho$.
 In the BEC regime one performs Bogoliubov transformation
 \be
 a_{\bf k}=u_{\bf k}b_{\bf k}+v_{\bf k}\bkrest_{\bf -k},\quad 
 \akrest_{\bf k}=u_{\bf k} \bkrest_{\bf k}+v_{\bf k} b_{\bf -k},
 \ee
with $[b_{\bf k} , \bkrest_{\bf p}]=\delta_{{\bf k},{\bf p}}$,
$\langle\bkrest_{\bf k}\bkrest_{\bf -k}\rangle=\langle b_{\bf k}b_{\bf -k} \rangle=0.$
As a result, the grand Hamiltonian is transformed to the 
Bogoluibov form:
\be
\ba
H=\widetilde{H}_0+\summa_{\bf k} E_{\bf k} \bkrest_{\bf k} b_{\bf k}+\dsfrac{1}{2}\summa_{\bf k}(E_{\bf k}-\omega_{\bf k}),
\lab{Eq26}
\ea
\ee
where $\langle\bkrest_{\bf k}b_{\bf k}\rangle = n_B(E_{\bf k},T)=1/[\exp (E_{\bf k}/T)-1]$ with
the phonon Goldstone mode dispersion  $E_{\bf k}^{2}=\omega_{\bf k}^2-X_1^2/4$. At small
momenta, this mode is a collective excitation of the condensate carrying spin $S_z=-1$,
while at large momenta it becomes the bare triplon mode. 

In accordance with Hugenholtz-Pines  theorem
\ci{Dickhoffbook} at
small $k$ the phonon dispersion is linear: $E_{\bf k}\sim ck+O(k^2)$, where $c$ can be interpreted as the 
speed of sound. This is achieved by setting $\omega_{\bf k}-X_1/2=\varepsilon_{\bf k}$,
that is, by:
\be
\mu-g\rhozero-2g\rhoone+g\sigmaone=0.
\lab{eq28}
\ee
This choice yields $E_{\bf k}=\sqrt{\veps_{\bf k}(\veps_{\bf k}+X_1)}$ with 
$c=\sqrt{X_1/2m}$, where $m$ is the triplon effective
mass. It can be shown  \ci{Andersen04, Rakhimov08} that
$X_1$ is related to the normal and
anomalous self energies as $\Sigma_{\rm n}=X_1/2+\mu$ and
 $\Sigma_{\rm a}=X_1/2$, respectively. The quantity $X_1$ 
plays a special role in our analysis: when  $X_1>0$,
the condensate is stable, otherwise it decays due to triplon-triplon interaction.
\cite{booknikinu,Ma71,Chung08}
Below we find $X_1$ in the $T-H_{\rm ext}$ plane and determine the stable BEC 
region by the condition $X_1\geq0$.

Using the explicit $u_{\bf k}=\sqrt{(\omega_{\bf k}+E_{\bf k})/2E_{\bf k}}$
 and $v_{\bf k}=\sqrt{(\omega_{\bf k}-E_{\bf k})/2E_{\bf k}}$, one obtains:
\begin{eqnarray}
&&\rho_1=\summa_{\bf k}\langle \akrest_{\bf k} a_{\bf k} \rangle=
\summa_{\bf k}\left(\dsfrac{\omega_{\bf k} W_{\bf k}}{E_{\bf k}}
-\dsfrac{1}{2}\right),  \nonumber \\
&&\sigma_1=\summa_{\bf k}\langle a_{\bf k} a_{\bf -k} \rangle=
2\summa_{\bf k} u_{\bf k}v_{\bf k}W_{\bf k} =
-\dsfrac{X_1}{2}\summa_{\bf k} \dsfrac{ W_{\bf k}}{E_{\bf k}},
\lab{eq29_1}
\end{eqnarray}
where $W_{\bf k}=n_B(E_{\bf k},T)+1/2$. 
Near the transition, $T\rightarrow T_{\rm c}$ the condensate fraction
vanishes: $\rho_0\rightarrow 0$, and Eq.\re{eq24} yields $X_1=0$. 
In the triplon BEC, the critical density  $\rho_c$ 
corresponds to $\mu_{\rm eff}=0$, i.e. $\rho_c=\mu/2g$.
Therefore, at a given chemical potential
$\mu=\widetilde{g}\mu_B H_{\rm ext}-\Delta$, where $\widetilde{g}$ is the 
electron Land\'e-factor, $T_{\rm c}$
is determined by: $\Sigma_{\bf k}n_B(\veps_{\bf k},T_{\rm c})={\mu}/{2g}.$

To perform MFA calculations one starts by solving Eqs.\re{eq24} and \re{eq28}
with $\rho_1$ and $\sigma_1$ given by Eq.\re{eq29_1}.
 In contrast to the BEC of atomic gases, in the triplon problem,
 the chemical potential is the input parameter,
 whereas the densities are the output ones.
 Bearing this in mind, we rewrite Eqs.\re{eq24} and \re{eq28} as:
\be
 X_1=2\mu+4g(\sigmaone-\rho_1),\qquad
 \rho_0=\dsfrac{X_1}{2g}-\sigma_1.
 \lab{eq211}
\ee
Using dimensional regularization at $T=0$, we can find from \re{eq29_1} 
more explicit expressions for the densities
\be
\ba
\rho_1=\rho_1(T=0)+\dsint\dsfrac{d^3 k}{(2\pi)^3}
n_B(E_{\bf k},T)\dsfrac{\veps_{\bf k}+X_1/2}{E_{\bf k}},\\
\sigma_1=\sigma_1(T=0)-\dsint\dsfrac{d^3 k}{(2\pi)^3}
n_B(E_{\bf k},T)\dsfrac{{X_1}/{2}}{E_{\bf k}},
\lab{eq212}
\ea
\ee
where $\sigma_1(T=0)=3\rho_1(T=0)=\sqrt{2}(m X_1)^{3/2}/4\pi^2$,
as shown in Ref.[\onlinecite{yukalov}]. 
By setting in all above formulas $\sigma_1=0$, one arrives at the HFP
approximation, \cite{Nikuni00,Misguich04} and particularly
\be
X_1^{\rm [HFP]}=2\mu-4g\rho_1,\qquad
\rho_0=\dsfrac{X_1^{\rm [HFP]}}{2g}.
\lab{eq213}
\ee
The above Eqs.\re{eq211}-\re{eq213} can be applied for any realistic $\veps_{\bf k}$.
It is instructive to note that for the parabolic dispersion
$\veps_{\bf k}={\bf k}^2/2m$, the BEC can be fully
described by only two parameters $\eta\equiv\mu m^3g^2$ and $t\equiv T/T_{\rm c}$ with
$T_{\rm c}={\widetilde{c}}\left({\mu}/{g}\right)^{2/3}/{m}$, 
$\widetilde{c}=\pi\left({\sqrt{2}}/{g_{3/2}(1)}\right)^{2/3}=2.0867$,
where $g_{3/2}(z)$ is the Bose function.\cite{Huangbook} 
The parameter $\eta$ is an analogue
of the gas parameter \cite{Huangbook} of atomic BEC.

\begin{figure}[tbp]
\epsfxsize=5cm \vspace{0.0cm}\epsfbox{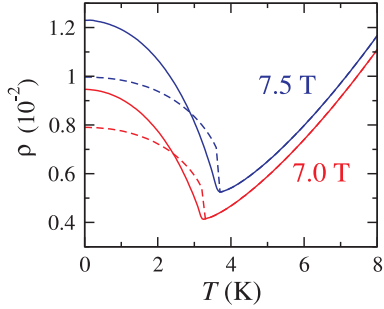} \vspace{0.0cm}
\caption{(Color online). Comparison of the HFB (solid) and the HFP
(dashed lines) results for the triplon density.
The HFB approach shows a continuous behavior, which fully agrees with the experimental
data \cite{Yamada08,Amore08}
while the HFP approach leads to the discontinuity.
The corresponding $H_{\rm ext}$ are marked near the plots.}
\end{figure}

Since the  MFA (both HFB and HFP) calculations 
are based on Eqs.\re{eq211}, \re{eq212}, 
a question about the existence of positive solutions for $X_1$ arises.
To analyze qualitatively the existence of the physical solutions,
we consider $T=0$ case. Here, the HFP Eq.\re{eq213} is simplified by substitution 
$Z_{\rm HFP}\equiv X_1^{\rm [HFP]}/2\mu$ to
$1=Z_{\rm HFP}+{2Z_{\rm HFP}^{3/2}\sqrt{\eta}}/{3\pi^2}$ and has physical solutions $Z_{\rm HFP}>0$
for any $\eta>0$.  This remains valid for all $t\le1$ at
any concentration $\rho$. However, in the HFB approximation the situation is different:
even at $t\le1$, the physical solutions of Eq.\re{eq211}  can disappear if 
$\eta$ exceeds a critical value $\eta_c$. For example, 
at $T=0$, Eq. \re{eq211} for $Z\equiv X_1/2\mu$ simplifies as 
\be 
1=Z-\dsfrac{4Z^{3/2}\sqrt{\eta}}{3\pi^2}.
\lab{eq217}
\ee

\begin{figure}[tbp]
\epsfxsize=5cm  \vspace{0.0cm}\epsfbox{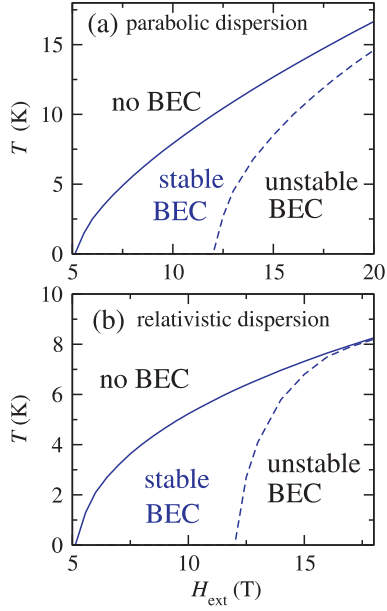} 
\caption{(Color online). Phase diagram for the parabolic (a) and relativistic (b) triplon dispersion
 in the HFB approximation.}
\end{figure}

When $\eta$ exceeds $\eta_c=\pi^4/12$, the  rhs in Eq.\re{eq217}
is less than 1 for any $Z\ge0$, therefore, it has no positive solutions,
and, as a result, $X_1$ acquires an imaginary part. 
Bearing in mind that $\eta=\mu m^3 g^2=(\mu_B\widetilde{g}H_{\rm ext}-\Delta)m^3 g^2$,
one concludes that even at $T=0$, if the  $H_{\rm ext}$
is strong enough the speed of sound $c=\sqrt{Z\mu/m}$ becomes complex and, hence, the BEC
is unstable. 

To calculate $\rho$ and $X_1$ one needs the bare $\varepsilon_{\bf k}$.
Misguich and Oshikawa \ci{Misguich04} demonstrated that only with the 
exact $\varepsilon_{\bf k}$ one can explain the overall $\rho_c-T_{\rm c}$ - dependence. 
Here we apply a similar approach, using a simpler,
"relativistic" $\veps_{\bf k}=\sqrt{\Delta^2+J^2k^2/4}-\Delta$, generic
for systems with gapped spectrum. This choice
leads to  $\rho_c\sim T_{\rm c}^{2}$ 
at higher and $\rho_c\sim T_{\rm c}^{3/2}$ at lower $T$s, respectively.\cite{Sherman03,Grether07}
Here the effective exchange $J=2\sqrt{\Delta/m}$ 
is chosen to match the parabolic and the relativistic $\varepsilon_{\bf k}$ at small $k$.

In numerical calculations we used parameters  by Yamada {\it et al.} \ci{Yamada08}
for TlCuCl$_3$: $m=0.0204$ K$^{-1}$, (i.e. $m=0.261\times10^{-25}$ g), unit cell size $0.79$ nm,
$\Delta=7.1$ K, $g=313$ K and $\widetilde{g}=2.06$. We neglect a weak renormalization
of the model parameters by temperature-dependent many-body effects, which can slightly
shift the stability region boundary, since we consider the regime of low $T$ and $\rho$.
This assumption yields a perfect agreement of theory and experiment \cite{Misguich04} in a similar
range of $T$ and $\rho$.
We begin with the comparison of the HFB and HFP approaches for the 
density $\rho$ in a constant $H_{\rm ext}$. Fig.1  shows a continuous plot of  
$\rho(T,H_{\rm ext})$ obtained with the HFB approach,\cite{Sirker04}  in full agreement with the experiment \ci{Yamada08,Amore08} and
in contrast to the HFP approximation. 
In Fig.2 we present the phase diagram obtained in the HFB for the parabolic
and the relativistic $\varepsilon_{\bf k}$. 
Solid curves in these figures 
present $T_{\rm c}$ vs $H_{\rm ext}$ obtained from $\Sigma_{\bf k}n_B(\veps_{\bf k},T_{\rm c})={\mu}/{2g}.$
The dashed lines present the BEC stability boundary: there is no $X_1>0$
solutions to the gap equations in the regions below these lines. 
Therefore, the HFB approach predicts 
the existence of a stable (the region between
solid and dashed lines) and unstable BEC zones (the region below the dashed line).
As expected, at low $T$ and small $H_{\rm ext}$ 
the stability region in Figs. 2(a) and 2(b) is the same for both $\varepsilon_{\bf k}$.
In general, the relativistic dispersion leads to a narrower stability zone
than the parabolic one. Note that magnetization measurements 
on TlCuCl$_3$ have been done for $H_{\rm ext}$ between 5.1 and 9 T.\cite{Yamada08,Amore08}
It would be interesting to experimentally study its behavior 
at higher $H_{\rm ext}$ to explore the instability region.\cite{Tanaka01} A direct access
to the dispersion and damping of the phonon-like mode in TlCuCl$_3$
can be achieved in the inelastic neutron scattering measurements.\cite{Ruegg}

\begin{figure}[tbp]
\epsfxsize=5cm  \vspace{0.0cm}\epsfbox{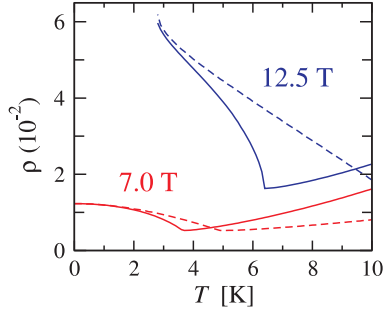} \vspace{0.0cm}
\caption{(Color online). Triplon density as a function of $T$ for relativistic (solid) and parabolic
(dashed lines) dispersion in the HFB approximation for $H_{\rm ext}$ marked near the plots.
The plot for $H=12.5$ T (upper curves) shows two anomalies, one of them
caused by the instability.}
\end{figure}

Density $\rho$ as a function of temperature is  presented in Fig.3
for two  $H_{\rm ext}$. At relatively weak fields, e.g.  $H_{\rm ext}=7.0$ T
the magnetization  exhibits only one anomaly at $T=T_{\rm c}$ while
at stronger one $H_{\rm ext}=12.5$ T, two anomalies are present.
The minimum at the solid line at 6.2 K is the onset of the BEC, while the
anomaly at $T$ slightly less than 3 K is due to the condensate decay.  
Similar physical behavior can be seen in Fig.4, which shows the
BEC fraction $\rho_0/\rho \times 100 \% $.
This fraction is rather large ($\sim 95\%  $ for $H_{\rm ext}=7.5$ T
at $T=0$) and rapidly decreases with increasing the temperature. 
In both Figs.3 and 4 the curves for $H_{\rm ext}=12.5$ T  start at $T\approx 3$ K
since the BEC is unstable below this $T$.
However, Fig.4 shows that even close to this point the condensate 
fraction is approximately $70\%$, and, therefore, in the instability zone the condensate can exist
for a short time \cite{Schilling09}  determined by the imaginary part of the self energy $X_1$.
This regime will be considered in an extended paper.

In summary, we have theoretically established the phase diagram 
of the field-induced triplon BEC in quantum antiferromagnets in the $T-H_{\rm ext}$ 
plane for a model relevant for the TlCuCl$_3$
compound. Our approach is based on the HFB approximation taking into account
the anomalous density in the condensate phase. We have shown that 
(i) at the BEC transition the magnetization remains
continuous demonstrating a minimum, in agreement with the experiment, 
(ii) in high magnetic fields the condensate becomes
unstable due to the triplon-triplon repulsion, resulting in interaction
of quasiparticles, and found the stability boundaries.  
The non-parabolic dispersion of triplons determined by the
crystal structure has the crucial effect on the phase diagram by
changing the boundaries $H_{\rm ext}^{(1)}$ and $H_{\rm ext}^{(2)}$
and making the stability region smaller.

\begin{figure}[tbp]
\epsfxsize=5cm \vspace{0.0cm}\epsfbox{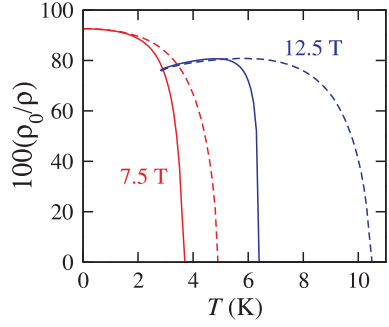}
\caption{(Color online). Condensate density fraction for relativistic 
(solid line) and parabolic (dashed line)
$\varepsilon_{\bf k}$ in the HFB approximation for $H_{\rm ext}$  marked near the plots.}
\end{figure}

We acknowledge support of the Volkswagen Foundation (AR)
and the University of the Basque Country UPV-EHU
Grant GIU07/40 (EYS). AR acknowledges support from Korea Science and Engineering
Foundation for visiting fellowship to Yonsei University. We are grateful to H. Kleinert, A. Pelster,
and O. Tchernyshyov for valuable discussions.

\bb{99}

\bi{Snoke07} R. Balili {\it et al.},
             Science {\bf 316}, 1007 (2007), 
             C. W. Lai {\it et al.},
             Nature  {\bf 417}, 47 (2002), 
             J. Keeling,
             Phys. Rev. B {\bf 74}, 155325 (2006) 	

\bi{Matsuda} T. Matsubara and H. Matsuda, Prog. Theor. Phys. {\bf 16}, 569 (1956); 
            H. Matsuda and T. Tsuneto, ibid {\bf 46}, 411 (1970); 

\bi{Tachiki}  M. Tachiki and T. Yamada, 
              J. Phys. Soc. Jpn. {\bf 28}, 1413 (1970)

\bi{Batyev} E.G. Batyev and S.L. Braginskii, JETP {\bf 60}, 781 (1984); 
            E.G. Batyev, {\it ibid} {\bf 62}, 173 (1985) 

\bi{Affleck} I. Affleck, Phys.Rev B {\bf 43}, 3215 (1991)

\bi{Giamarchi99} T. Giamarchi and A. M. Tsvelik,
		 Phys. Rev. B {\bf 59}, 11398 (1999)

\bi{Nikuni00}  T. Nikuni {\it et al.},
               Phys. Rev. Lett. {\bf 84}, 5868 (2000)

\bi{Crisan05}  M. Crisan {\it et al.},
               Phys. Rev. B {\bf 72}, 184414 (2005)
		 
\bi{Stone07} M. B. Stone {\it et al.},
             New J. Phys. {\bf 9},  31 (2007)

\bi{Ruegg07} Ch. C. R\"{u}egg {\it et al.},
             Phys. Rev. Lett. {\bf 98}, 017202 (2007) 

\bi{Yamada08} F. Yamada {\it et al.},
              J. Phys. Soc. Jpn. {\bf 77}  013701 (2008)

\bi{Amore08} R. Dell'Amore, A. Schilling, and K. Kr\"{a}mer
             Phys. Rev. B {\bf 79} 014438 (2009);
             R. Dell'Amore, A. Schilling, and K. Kr\"{a}mer, 
             Phys. Rev. B {\bf 78} 224403 (2008)

\bi{Aczel09} A. A. Aczel {\it et al.},
	     Phys. Rev. B {\bf 79}, 100409(R) (2009)
	
\bi{Paduan09}  A. Paduan-Filho {\it et al.},
	       Phys. Rev. Lett. {\bf 102}, 077204 (2009)
	
\bi{Laflorencie09} N. Laflorencie and F. Mila,
                   Phys. Rev. Lett. {\bf 102}, 060602 (2009)

\bi{Demidov08} For the BEC in magnetic systems under external pumping:
               V. E. Demidov {\it et al.},
               Phys. Rev. Lett. {\bf 100}, 047205 (2008), 
               for theory:
               Yu. D. Kalafati and V. L. Safonov, JETP Lett. {\bf 50}, 149 (1989); 
               I. S. Tupitsyn, P. C. Stamp, and A. L. Burin,
               Phys. Rev. Lett. {\bf 100}, 257202 (2008);
               A. I. Bugrij and V. M. Loktev, Low Temp. Phys. {\bf 33}, 37 (2007).

\bi{Giamarchi08} T. Giamarchi, C. R\"{u}egg, and  O. Tchernyshyov,
                 Nature Physics {\bf 4}, 198 (2008)

\bi{Huangbook}   Kerson Huang, 
                {\it Statistical Mechanics}
                 Wiley (1987)

\bi{Rakhimov08}        A. Rakhimov {\it et al.},
                       Phys. Rev. A {\bf 77}, 033626 (2008)

\bi{Misguich04} G. Misguich and M. Oshikawa,
                J. Phys. Soc. Jpn. {\bf 73}, 3429 (2004)

\bi{Andersen04}      J. O. Andersen,
                     Rev. Mod. Phys. {\bf 76}, 599 (2004)

\bi{yukalov} V. I. Yukalov and H. Kleinert, 
            \pra {\bf 73}, 063612, (2006); 
            V.I. Yukalov, Ann. Phys. {\bf 323} 461 (2008)             
	     
\bi{Dickhoffbook} W. H. Dickhoff and D. Van Neck, 
                 {\it Many-Body Theory Exposed}
	          World Scientific (2005)

\bi{booknikinu}  A. Griffin,  T. Nikuni and E. Zaremba, 
                {\it Bose-Condensed Gases at Finite Temperatures}
                 Cambridge University Press (2009)

\bi{Ma71} S.-K. Ma, H. Gould, and V. K. Wong, 
         Phys. Rev. A {\bf 3}, 1453 (1971)
		 
\bi{Chung08}	M.-C. Chung and A. B. Bhattacherjee, 
	        New Journ. of Phys. {\bf 11}, 123012 (2009) 

\bi{Sherman03} E. Ya. Sherman {\it et al.},
               Phys. Rev. Lett. {\bf 91}, 057201 (2003)

\bi{Grether07} Theory of the BEC in the ideal relativistic Bose
               gas was developed in 
               M. Grether, M. de Llano, and George A. Baker, Jr.,
	       Phys. Rev. Lett. {\bf 99}, 200406 (2007)

\bi{Sirker04}    The continuity can be achieved 
                 in the HFP approach with a strong Dzyaloshinskii-Moriya interaction: 
                 J. Sirker, A. Wei{\ss}e, O.P. Sushkov,
	         Europhys. Lett. {\bf 68}, 275 (2004)

\bi{Tanaka01} Neutron scattering experiments by 
              H. Tanaka {\it et al.},  J. Phys. Soc. Jpn. {\bf 70}, 939 (2001) 
              explored $H_{\rm ext}$ up to 12 T at $T=1.9$ K, lower than 
              calculated $H_{\rm ext}^{(2)}$. To observe
              unambiguously the effects of instability, experiment has to be performed 
              deeply in the instability region.

\bi{Ruegg} Ch. R\"{u}egg {\it et al.},
           Nature {\bf 423}, 62 (2003) 

\bi{Schilling09} For other aspects of the BEC lifetime see 
                 A. Schilling, arXiv:0908.3033v1

\eb
\edc